# Services in Android can Share Your Personal Information in Background

Manoj Kumar[1] and Sheshendra Rathi[2]

[1]Department of Computer Science, Maharaja Surajmal Institute, New Delhi, India

[2]Eriwssel Network Private Limited

*Abstract*— Mobile phones have travelled a very long journey in a very short span of time since its inception in 1973. This wonderful toy of 20th century has started playing significant role in people's life. More than 5 billion mobile users are there around the world and almost 90% of the entire earth is under the mobile coverage now. Today's smart phones are equipped with numerous features, faster processors and high storage capacity. Android is a latest trend in this series whose popularity is growing by leaps and bounds. Android has a number of components which helps Application developers to embed distinguish features in applications. This paper explains how the Service component of Android can share your personal information to others without users' interaction.

## I. INTRODUCTION

Mobile phones, Smart phones and faster web browsers are changing the interface to the clients. The cloud based service which was a dream few years ago is available on mobile web due to the increased bandwidth and connectivity. HTML5 adds many new features and streamlines functionality in order to render processor intensive add-ons unnecessary for many common functions. It will be a particular boon to those smart phones for which supporting Flash has been problematic.

With so many platforms available today, Android seems to be most promising. Android is a software stack for mobile devices that includes an operating system, middleware and key Applications [1]. Android has been making waves as the best smart phone platform since its inception in 2007. Android platform market has increased tremendously with its popularity. ANALYST OUTFIT Canalys claims that Android will continue to grow at more than twice the rate of its major smart phone competitors in 2011 [2]. Gingerbread, the eighth update of Google's Android is there in the market. Whereas, latest released Honeycomb version supports multicore-processors and hardware acceleration for graphics. Android has been activated over 100 million devices worldwide. With 36 OEM's 215 carriers, 310 devices and 112 countries, Android has become a leading platform among others. 4.5 billion Apps have been installed so far with 200 K Apps available in wide variety of categories in Android market.

## II. FEATURES OF ANDROID

Developing Apps for mobile phones is a different experience than developing desktop applications, web-applications or back-end server processes. The basic components of Android are similar but they are packaged differently to make phones more crash-resistant. The main components of the Android framework are:

*A. Activities*

An Activity represents a single screen with a user interface [3]. An Activity can be considered as Android analogue for the window or dialog box in a desktop application.

*B. Content Providers*

A content provider manages a shared set of data [3]. The Android framework encourages you to make your data available to other applications.

*C. Broadcast Receivers*

A broadcast receiver is a component that responds to system-wide broadcast announcements [3]. It notifies applications of various events from hardware state changes (e.g. an SD card was inserted), to incoming data (e.g. an SMS has arrived).

*D. Services*

A service is a component which runs in the background, without interacting with the user. At number of occasions, applications will need to run processes for a long time without any interventions from the users. Activities and content providers are short-lived and can be shut down at any time. Services on the other hand, are designed to keep running.

## III. LIFE CYCLE OF A SERVICE COMPONENT

An Activity starts and stops a service to do some work for it in the background, examples include playing music even if the player activity gets garbage collected, polling the internet for RSS feed/Atom feed updates and maintaining an online chat connection even if the client loses focus due to an incoming phone call [7]. Any service is started by calling startService function. Later onCreate and onStart functions are called to create and start the service respectively. At the end on Destroy function is called to terminate the service. Figure 1 explains the life cycle of the service component.

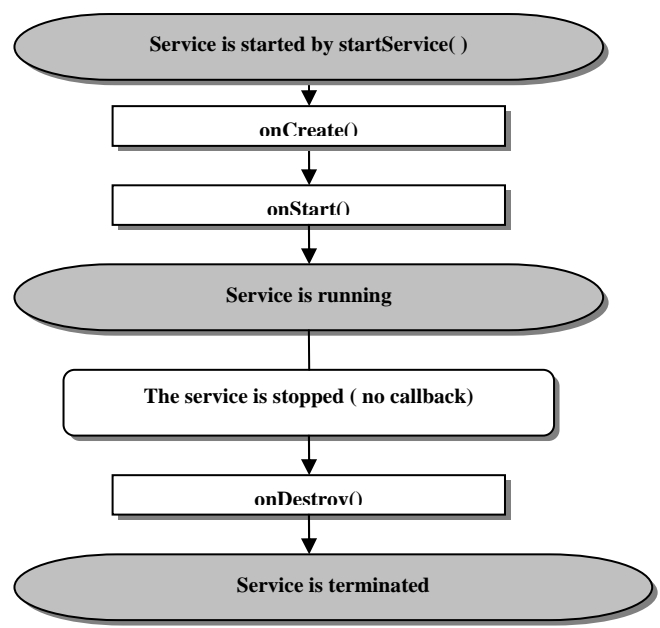

Figure 1: Lifecycle of service

## IV. IMPLEMENTATION OF SERVICE COMPONENT

### A. Folders and Files in Android Framework

While developing any App on Android platform, src and res folders with other folders are created. There will be .java file in src folder which is responsible for the main activity. Activity is the first building block through which user interacts with the App. We can create other classes in the same src folder. In res folder we have drawable, layout and values folders. All the layouts are designed through main.xml in layout folder. AndroidManifest.xml file is used to contain the essential information about the application. It declares the minimum level of the API required for the application. With several other information's, this file contains the necessary permission to run the application. AndroidManifest file is also used to keep a track on the different versions of the same App by the Android market.

### B. Coding

Two classes namely ServiceDemo and MyService have been defined in the following way

#### 1) ServiceDemo.java

```
1. package com.example.ServiceDemo;

/****************************************************
****************************************************
Declaration of all the import directives for the required
packages
****************************************************
****************************************************/

2. public class ServiceDemo extends Activity
   implements OnClickListener{
3. private static final String TAG="ServiceDemo";
4. Button buttonStart,buttonStop;

/****************************************************
****************************************************
      P U B L I C   F U N C T I O N S
****************************************************
****************************************************/

       // 1. Called when the activity is first created

  @Override
5. public void onCreate(Bundle savedInstanceState)
   {
6. super.onCreate(savedInstanceState);
7. setContentView(R.layout.main);

       // 2. Two buttons buttonStart and buttonStop are
       //    there in main.xml

8. buttonStart = (Button)findViewById(R.id.buttonStart);

9. buttonStop = (Button)findViewById(R.id.buttonStop);

       // 3. Attaching onClickListener with both
       //    the buttons

11. buttonStart.setOnClickListener(this);

12. buttonStop.setOnClickListener(this);
    }

  @Override

13. public void onClick(View src) {

          // 4. TODO Auto-generated method stub

14. switch(src.getId())
    {
15. case R.id.buttonStart:
16. Log.d(TAG, "onClick: starting srvice");

        // 5. Starting the service with the click of
        //    start button

17. startService(new Intent(this, MyService.class));
18. break;
19. case R.id.buttonStop:
20. Log.d (TAG, "onClick: stopping srvice");

        // 6. Terminating the service with the
        //    click of start button

21. stopService(new Intent(this, MyService.class));
22. break;
     }
    }
  }
```

To demonstrate the effect of service component, an activity is being created. As the layout contains start and stop buttons (Figure 2), buttonStart and buttonStop have been declared correspondingly in the ServiceDemo class (line 4). Declared variables have been attached with the buttons created in layout using setOnClickListener method ( line 11, 12). Further, inside the function onClick two cases have been defined. When the user clicks on start button, it makes an entry in log file (line 17), it will start the service in background by calling startService function (line18). In the next case, when the user clicks on stop button, it will again make an entry in log file (line 21). It finally stops the service by calling stopService function (line 22). startService and stopService methods call the other methods of MyService class.

TelephonyManager class provides access to information about the telephony service on the device. Applications can use the methods in this class to determine telephony services and states as well as to access some type of subscriber information [3].To retrieve the subscriber's personal information the necessary permission must be taken through AndroidManifest.xml file.

```
MyService.class
```

1. package com.example.ServiceDemo;

/***************************************************
  ***************************************************
  Declaration of all the import directives for the required packages

  ***************************************************
  ***************************************************/

2. public class MyService extends Service{

/***************************************************
  ***************************************************
        P R I V A T E   V A R I A B L E S
  ***************************************************
  ***************************************************/

3. private static final String TAG = "MyService";
4. private String imei;
5. private String msisdn;
6. private String nwOp;
7. private int nwtype;
8. private String country;
9. private String version;
10. private int phtype;
11. private long time;
12. private Date date;
    @Override
13. public IBinder onBind(Intent intent) {

        // 1. TODO Auto-generated method stub

14. return null;

        // 2. onCreate( ) Method will be called at the
              creation of the service
    @Override
15. public void onCreate() {
16. Toast.makeText(this, "My Service Created",
    Toast.LENGTH_LONG).show();
17. Log.d (TAG, "onCreate");
                                }
    // 3. onStart() Method will be called at the
    //    starting of the service
    @Override
18. public void onStart(Intent intent, int startid) {
19. Toast.*makeText*(this, "My Service Started",
    Toast.*LENGTH_LONG*).show();
20. try
        {
        // 4. Retrieve Phone Related Information
21. TelephonyManager mTelephonyMgr =
    (TelephonyManager)getSystemService(
    *TELEPHONY_SERVICE*);
22. imei = mTelephonyMgr.getDeviceId();
23. msisdn = mTelephonyMgr.getLine1Number();
24. nwOp =
    mTelephonyMgr.getNetworkOperatorName();
25. nwtype = mTelephonyMgr.getNetworkType();

26. country =
    mTelephonyMgr.getNetworkCountryIso();
27. version =
    mTelephonyMgr.getDeviceSoftwareVersion();
28. phtype = mTelephonyMgr.getPhoneType();

        // 5. Set the Criteria parameters
29. Criteria cr = new Criteria();
30. cr.setAccuracy(2);

        // 6. Date and time
31. date = new Date();
32. time = date.getTime();
33. Toast.*makeText*(this," all the values hav been
    read successfully",Toast.*LENGTH_LONG*).
    show();
        }
34. catch(Exception E)
        {
35. Log.*d*(*TAG*, "Error in reading variables in
    onStart");
        }
36. Log.*d*(*TAG*, "onStart");
    }
        // 7. onDestroy() Method will be called at the
end of the service
    @Override
37. public void onDestroy() {
38. try {
39. Log.*i*(*TAG*, "values of the different variables");

        // 8. Values of the different variables which
were
        //     collected in the method onStart() are
being put
        //     in the log before the service is going to
        //     terminate
40. Log.i("imei",imei);
41. Log.i("network operator",nwOp);
42. Log.i("network type", Integer.toString
    (nwtype));
43. Log.i("country",country);
44. Log.i("phtype",Integer.toString(phtype));
45. Log.i("time",Long.toString(time));
46. Log.i("date",date.toString());
47. Log.i("version",version);
48. Log.i("Phone number",msisdn);
        }
49. catch (Exception e) {
        // 9. TODO Auto-generated catch block
50. Log.*i*(*TAG*, "error in writing different variables
    to logcat");
        }

51. Toast.*makeText*(this, "My Service Stopped,
    Check Log now", Toast.*LENGTH_LONG*).
    show();
52. Log.*d*(*TAG*, "onDestroy");
                                }
    }

To understand the execution of service component, different varaibles have been declared in MyService.class. Other variables may also be declared as per the need. MyService.class explains that how the vital information of a mobile user collected without informing the user. onCreate, onStart and onDestroy are the 3 methods, which governs the entire lifecycle of service utility. onCreate and onStart functions are called from onClick method of ServiceDemo class, when the user clicks the start button. Inside the onStart method, values in the declared variables are collected by calling appropriate functions.

Finally, when the user clicks the stop button, onDestroy method is called. Before the service terminates, the collected values of the varaibles are put in the log file. The collected information may either be saved in a databse file or can be updated to a server directly if the internet connectivity is available. In the entire process of collecting and storing the values of different variable, no interaction was done with the user.

## V. OUTPUT

Figure 2 displays the activity screen on an emulator with start and stop button. Service utility starts with a click on start button and will stop with a click on stop button. All the processing will be done in the background.

Figure 3 displays the logcat file. However several options are available, logcat file has been chosen to keep the values of the different variable read from the mobile phone. Serial numbers have been included with some of the lines in the logcat file which are concerned to our purpose. Serial number 1 and 2 shows the creating and starting of the service. Serial number 3 shows the stopping of the service. Serial number 5 to 13 shows those values which have been read from the mobile phone.

## VI. CONCLUSION

This paper explains the way in which the information of any of the mobile user can be collected without much interaction with the user. It can be understood by the entries of the different variables in the logcat file that how easily the privacy of a mobile user can be compromised. SQLite database can be used and values of the variables can be stored in a file. The file can be sent to a server dedicated to collect such files when the internet connection is available. However on the other end, the service component of Android can be useful in other instances. A number of apps require the service utility.

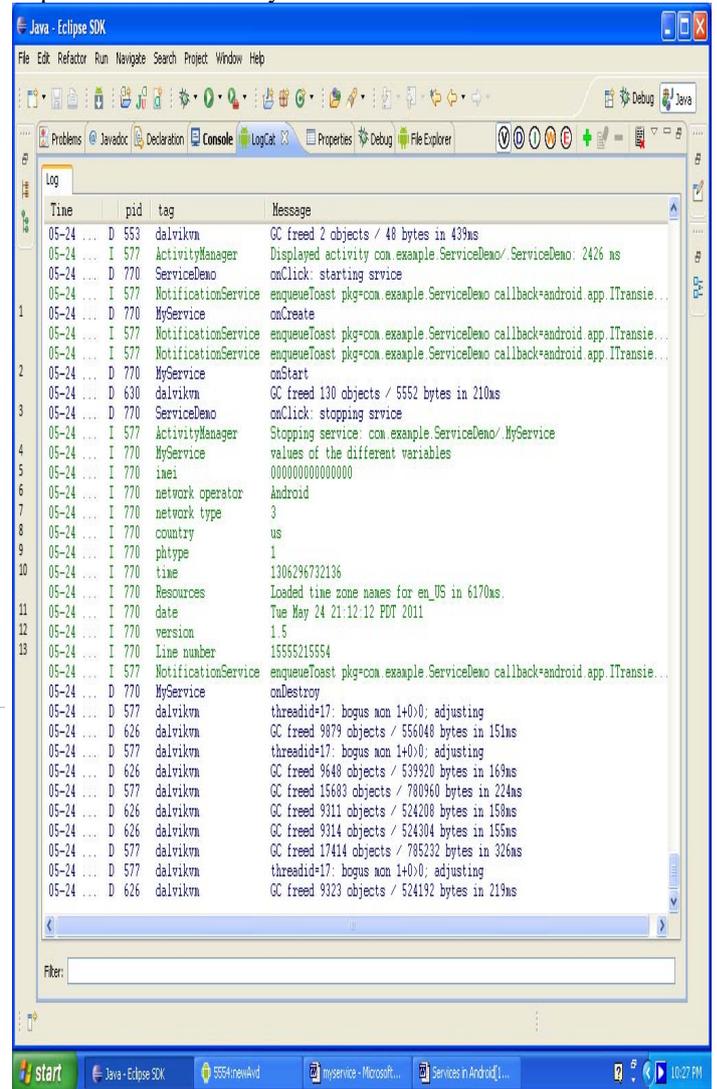

Fig. 3 Logcat file

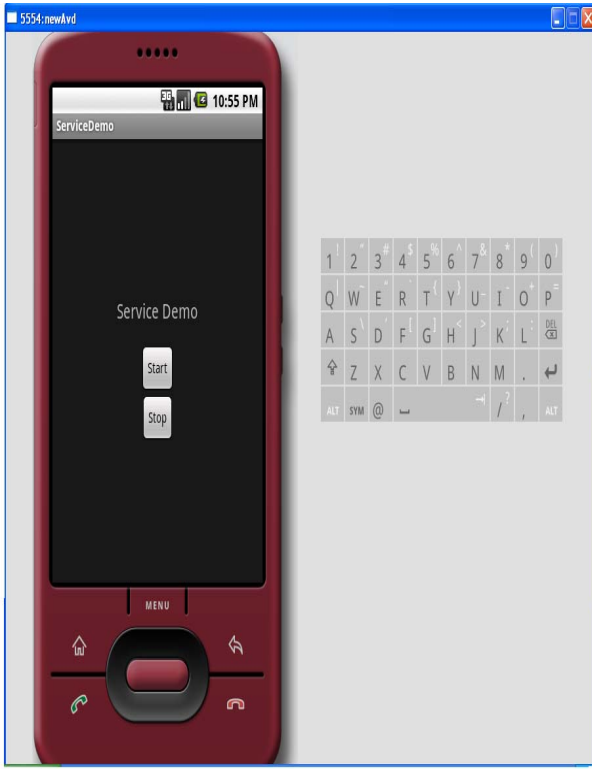

Fig. 2: Emulator screenshots